\documentstyle[12pt,epsfig]{article}
\begin{document}

\vspace{0.2cm}

\begin{center}
{\large \bf The Glashow resonance as a discriminator of UHE cosmic
neutrinos originating from $p\gamma$ and $pp$ collisions}
\end{center}

\vspace{0.2cm}
\begin{center}
{\bf Zhi-zhong Xing} \footnote{E-mail: xingzz@ihep.ac.cn} \\
{\sl Institute of High Energy Physics, Chinese Academy of
Sciences, Beijing 100049, China \\
and Center for High Energy Physics, Peking University,
Beijing 100080, China} \\
\vspace{0.7cm}
{\bf Shun Zhou} \footnote{E-mail: zhoush@mppmu.mpg.de} \\
{\sl Max-Planck-Institut f\"{u}r Physik
(Werner-Heisenberg-Institut), 80805 M\"{u}nchen, Germany}
\end{center}

\vspace{2.0cm}

\begin{abstract}
We re-examine the interesting possibility of utilizing the Glashow
resonance (GR) channel $\overline{\nu}^{}_e + e^- \to W^- \to {\rm
anything}$ to discriminate between the UHE cosmic neutrinos
originating from $p\gamma$ and $pp$ collisions in an optically thin
source of cosmic rays. We propose a general parametrization of the
initial neutrino flavor composition by allowing the ratios
$\Phi^{p\gamma}_{\pi^-}/\Phi^{p\gamma}_{\pi^+}$ and
$\Phi^{pp}_{\pi^-}/\Phi^{pp}_{\pi^+}$ to slightly deviate from their
conventional values. A relationship between the typical source
parameter $\kappa \equiv (\Phi^{p\gamma}_{\pi^+} +
\Phi^{p\gamma}_{\pi^-})/(\Phi^{pp}_{\pi^+} + \Phi^{pp}_{\pi^-} +
\Phi^{p\gamma}_{\pi^+} + \Phi^{p\gamma}_{\pi^-})$ and the working
observable of the GR $R^{}_0 \equiv \Phi^{\rm
T}_{\overline{\nu}^{}_e}/ (\Phi^{\rm T}_{\nu^{}_\mu} + \Phi^{\rm
T}_{\overline{\nu}^{}_\mu})$ at a neutrino telescope is derived, and
the numerical dependence of $R^{}_0$ on $\kappa$ is illustrated by
taking account of the latest experimental data on three neutrino
mixing angles. It is shown that a measurement of $R^{}_0$ is in
principle possible to identify the pure $p\gamma$ interaction
($\kappa =1$), the pure $pp$ interaction ($\kappa =0$) or a mixture
of both of them ($0 < \kappa < 1$) at a given source of UHE cosmic
neutrinos. The event rate of the GR signal against the background is
also estimated.
\end{abstract}
\begin{center}
\end{center}

\newpage

\section{Introduction}

The full construction of the IceCube detector \cite{IceCube}, a
${\rm km}^3$-scale neutrino telescope at the South Pole, has
recently been completed. It offers a great opportunity to discover
ultrahigh-energy (UHE) cosmic neutrinos, whose existence may
hopefully allow us to pin down the origin of UHE cosmic rays. The
reason is simply that the UHE cosmic protons originating in a cosmic
accelerator, such as a gamma ray burst or active galactic nuclei
\cite{Halzen}, unavoidably interact with ambient photons or protons.
Such energetic $pp$ or $p\gamma$ interactions produce a large amount
of charged pions, from which UHE cosmic neutrinos can copiously be
produced. Since UHE cosmic neutrinos are not deflected by the
interstellar magnetic field, they can be used to locate the cosmic
accelerators if they are observed in a terrestrial neutrino
telescope.

The $p\gamma$ and $pp$ collisions at an optically thin source of UHE
cosmic rays are usually referred to as the conventional production
mechanism of UHE cosmic neutrinos. Charged pions are mainly produced
via $p + \gamma \to \Delta^+ \to \pi^+ + n$ in the $p\gamma$
interaction or $p + p \to \pi^{\pm} + X$ with $X$ being other
particles in the $pp$ interaction \cite{XZ}. So neutrinos arise from
the decay chain $\pi^+ \to \mu^+ + \nu^{}_\mu \to e^+ + \nu_e +
\overline{\nu}^{}_\mu + \nu^{}_\mu$ and its charge-conjugate
process. In an astrophysical source of either $p\gamma$ or $pp$
collisions one has the same $\nu^{}_\alpha +
\overline{\nu}^{}_\alpha$ flavor distribution $\Phi^{\rm S}_e :
\Phi^{\rm S}_\mu : \Phi^{\rm S}_\tau = 1 : 2 : 0$, where $\Phi^{\rm
S}_\alpha \equiv \Phi^{\rm S}_{\nu_\alpha} + \Phi^{\rm
S}_{\overline{\nu}_\alpha}$ with $\Phi^{\rm S}_{\nu_\alpha}$ and
$\Phi^{\rm S}_{\overline{\nu}_\alpha}$ being the fluxes of
$\nu^{}_\alpha$ and $\overline{\nu}^{}_\alpha$ (for $\alpha = e,
\mu, \tau$) at the source. This initial flavor distribution is
expected to change to $\Phi^{\rm T}_e : \Phi^{\rm T}_\mu : \Phi^{\rm
T}_\tau = 1 : 1 : 1$ at a neutrino telescope such as the IceCube,
because UHE cosmic neutrinos may oscillate many times on the way to
the Earth and finally reach a flavor democracy \cite{Pakvasa} if the
$3\times 3$ neutrino mixing matrix $V$ satisfies the $|V^{}_{\mu i}|
= |V^{}_{\tau i}|$ condition (for $i=1,2,3$) \cite{XZ08}. Provided
such a flavor democracy is really measured at the IceCube detector
or at a more advanced neutrino telescope in the future, one will be
essentially convinced that the measured UHE cosmic neutrinos come
from the $p\gamma$ or $pp$ collisions (or a mixture of both of them)
in a distant cosmic accelerator. Then an immediate and meaningful
question is whether the neutrino telescope can discriminate between
the $p\gamma$ and $pp$ interactions at the source.

The answer to the above question is in principle affirmative, if the
$\nu^{}_e$ and $\overline{\nu}^{}_e$ fluxes can separately be
determined at a neutrino telescope. Unfortunately, the present
IceCube detector is unable to distinguish between the Cherenkov
light patterns arising from the interactions of $\nu^{}_e$ and
$\overline{\nu}^{}_e$ with ice. A possible way out is to detect the
UHE cosmic $\overline{\nu}^{}_e$ flux by means of the Glashow
resonance (GR) channel $\overline{\nu}^{}_e + e^- \to W^- \to {\rm
anything}$ \cite{Glashow,Berezinsky}, whose cross section can be
about two orders of magnitude larger than the cross sections of
$\overline{\nu}^{}_e N$ interactions around the resonant energy
$E^{}_{\overline{\nu}^{}_e} \simeq 6.3$ PeV \cite{Gandhi}. As
pointed out by Anchordoqui {\it et al} \cite{Weiler}, the GR may
serve for a useful discriminator of UHE cosmic neutrinos originating from
$p\gamma$ and $pp$ collisions in an optically thin source of cosmic
rays. The main purpose of the present paper is to re-examine this
interesting possibility by paying particular attention to the flavor
content of UHE cosmic neutrinos and its variation from a source to a
telescope.

Our work is different from the previous attempts in this connection
(e.g., Ref. \cite{Pakvasa} and Refs. \cite{Weiler}---\cite{Lin})
in several aspects.
First, we propose a general parametrization of the initial flavor
distribution of UHE cosmic neutrinos originating from $p\gamma$ and
$pp$ collisions by allowing
$\Phi^{p\gamma}_{\pi^-}/\Phi^{p\gamma}_{\pi^+} \neq 0$ and
$\Phi^{pp}_{\pi^-}/\Phi^{pp}_{\pi^+} \neq 1$. This treatment makes
sense as the assumptions $\Phi^{p\gamma}_{\pi^-} = 0$ (in the
$p\gamma$ interaction) and $\Phi^{pp}_{\pi^-} = \Phi^{pp}_{\pi^+}$
(in the $pp$ interaction) may not exactly hold in a realistic cosmic
accelerator. Second, we establish an analytical relationship between
three typical source parameters ($\delta^{}_{p\gamma} \equiv
\Phi^{p\gamma}_{\pi^-}/\Phi^{p\gamma}_{\pi^+}$, $\delta^{}_{pp}
\equiv \Phi^{pp}_{\pi^-}/\Phi^{pp}_{\pi^+} - 1$ and $\kappa \equiv
[\Phi^{p\gamma}_{\pi^+} + \Phi^{p\gamma}_{\pi^-}]/[\Phi^{pp}_{\pi^+}
+ \Phi^{pp}_{\pi^-} + \Phi^{p\gamma}_{\pi^+} +
\Phi^{p\gamma}_{\pi^-}]$) and the working observable of the GR
($R^{}_0 \equiv \Phi^{\rm T}_{\overline{\nu}^{}_e}/
[\Phi^{\rm T}_{\nu^{}_\mu} + \Phi^{\rm T}_{\overline{\nu}^{}_\mu}]$)
at a neutrino telescope
\footnote{Note that $X^{}_\gamma$ and $T$ have been used in
Ref. \cite{Winter1} to describe the fraction of UHE cosmic neutrinos
produced from the $p\gamma$ interaction and the working observable
at the neutrino telescope, respectively.}.
Third, we examine the numerical dependence of $R^{}_0$ on $\kappa$
by taking account of the latest experimental data on three neutrino
mixing angles. Our result shows that a measurement of $R^{}_0$ is in
principle possible to identify the pure $p\gamma$ interaction
($\kappa =1$), the pure $pp$ interaction ($\kappa =0$) or a mixture
of both of them ($0 < \kappa < 1$) at a given astrophysical source,
in particular after all the neutrino mixing parameters of $V$ are
well determined from a variety of terrestrial neutrino oscillation
experiments. In addition, the event rate of the GR signal against the
relevant background is also estimated in this paper.

\section{Modified Flavor Distribution on the GR}

We have denoted the $\pi^\pm$ fluxes from the $p\gamma$ interaction
as $\Phi^{p\gamma}_{\pi^\pm}$, and those from the $pp$ interaction
as $\Phi^{pp}_{\pi^\pm}$. In the conventional picture of $p\gamma$
collisions one mainly considers the $\Delta$-resonance channel $p +
\gamma \to \Delta^+ \to n + \pi^+$, and thus $\Phi^{p\gamma}_{\pi^-}
=0$ is taken as a good approximation for a given astrophysical
source. As for the $pp$ interaction in a cosmic accelerator, the
produced $\pi^+$, $\pi^-$ and $\pi^0$ mesons are expected to be in
almost equal amount due to the isospin symmetry. Hence
$\Phi^{pp}_{\pi^-} = \Phi^{pp}_{\pi^+}$ is also a good
approximation. In general, however, a small amount of $\pi^-$ mesons
should be produced from the $p\gamma$ interaction (e.g., from the
multi-pion production channel $p + \gamma \to n + \pi^+ + {\bf
n}(\pi^+\pi^-)$ with ${\bf n}$ being a positive integer
\cite{Glashow2}
\footnote{Note that the back reaction $n +
\gamma \to p + \pi^-$ could also produce $\pi^-$ mesons if the
optical thickness of the source is non-negligible, and the
$\overline{\nu}^{}_e$ flux originating from the
beta decays of neutrons might even dominate in some astrophysical
sources for very specific energy ranges \cite{Winter2}.
For simplicity, here we follow Ref. \cite{Weiler} and focus on the
cases in which the afore-mentioned effects can be neglected.}),
and a slight difference between $\Phi^{pp}_{\pi^-}$ and
$\Phi^{pp}_{\pi^+}$ must be present for the $pp$ interaction. So we
consider a general source in which both $p\gamma$ and $pp$
collisions are important. To be explicit, we define three typical
source parameters to describe the content of $\pi^+$ and $\pi^-$
mesons produced from $p\gamma$ and $pp$ collisions:
$\delta^{}_{p\gamma} \equiv
\Phi^{p\gamma}_{\pi^-}/\Phi^{p\gamma}_{\pi^+}$, $\delta^{}_{pp}
\equiv \Phi^{pp}_{\pi^-}/\Phi^{pp}_{\pi^+} - 1$ and
\begin{eqnarray}
\kappa \equiv \frac{\Phi^{p\gamma}_{\pi^+} +
\Phi^{p\gamma}_{\pi^-}}{\Phi^{pp}_{\pi^+} + \Phi^{pp}_{\pi^-} +
\Phi^{p\gamma}_{\pi^+} + \Phi^{p\gamma}_{\pi^-}} \; .
\end{eqnarray}
In this simple parametrization the $\kappa = 1$ and $\kappa =0$
cases correspond to the pure $p\gamma$ and pure $pp$ interactions,
respectively. If the value of $\kappa$ is found to lie in the $0 <
\kappa < 1$ range at a neutrino telescope, it will imply that both
$p\gamma$ and $pp$ collisions exist at the relevant astrophysical
source.

Now we look at the flavor composition of UHE cosmic neutrinos
originating from $p\gamma$ and $pp$ collisions in an optically thin
source of cosmic rays. Taking account of $\kappa$,
$\delta^{}_{p\gamma}$ and $\delta^{}_{pp}$ defined above, we obtain
the ratio of neutrino and antineutrino fluxes as follows:
\begin{eqnarray}
\hspace{-0.2cm} && \hspace{-0.2cm} \left\{{\Phi^{\rm S}_{\nu_e}:
\Phi^{\rm S}_{\overline\nu_e}: \Phi^{\rm S}_{\nu_\mu}: \Phi^{\rm
S}_{\overline\nu_\mu}:\Phi^{\rm
S}_{\nu_\tau}:\Phi^{\rm S}_{\overline\nu_\tau}}\right\} \nonumber \\
\hspace{-0.2cm} & = & \hspace{-0.2cm} \left(\Phi^{p\gamma}_{\pi^+} +
\Phi^{pp}_{\pi^+}\right) \left\{\frac{1}{3}: 0: \frac{1}{3}:
\frac{1}{3}: 0: 0\right\} +
\left(\Phi^{p\gamma}_{\pi^-} + \Phi^{pp}_{\pi^-}\right)
\left\{0: \frac{1}{3}: \frac{1}{3}: \frac{1}{3}: 0: 0\right\} \nonumber \\
\hspace{-0.2cm} &=& \hspace{-0.2cm}
\left\{\frac{1}{3} \left[\frac{1}{2+\delta^{}_{pp}} +
\frac{1+\delta^{}_{pp} -
\delta^{}_{p\gamma}}{(2+\delta^{}_{pp})
(1+\delta^{}_{p\gamma})}\kappa
\right]: \frac{1}{3} \left[\frac{1+\delta^{}_{pp}}{2+\delta^{}_{pp}} -
\frac{1+\delta^{}_{pp} -
\delta^{}_{p\gamma}}{(2+\delta^{}_{pp})
(1+\delta^{}_{p\gamma})} \kappa
\right]: \frac{1}{3}:\frac{1}{3}:0:0\right\} . \; ~~~~~
\end{eqnarray}
Given the definition $\Phi^{\rm S}_\alpha \equiv \Phi^{\rm
S}_{\nu_\alpha} + \Phi^{\rm S}_{\overline{\nu}_\alpha}$ (for $\alpha
= e, \mu, \tau$), it is straightforward to arrive at the
conventional $\nu^{}_\alpha + \overline{\nu}^{}_\alpha$ flavor
distribution $\Phi^{\rm S}_e : \Phi^{\rm S}_\mu : \Phi^{\rm S}_\tau
= 1 : 2 : 0$. This simple result is completely independent of three
source parameters. That is why one has to separately measure the
$\nu^{}_e$ and $\overline{\nu}^{}_e$ fluxes at a neutrino telescope
so as to probe $\Phi^{\rm S}_{\nu_e}$ and $\Phi^{\rm
S}_{\overline\nu_e}$ at the astrophysical source.

Thanks to the effect of neutrino oscillations, the $\nu^{}_\beta$
and $\overline{\nu}^{}_\beta$ fluxes observed at the telescope are
simply given by
\begin{eqnarray}
\Phi^{\rm T}_{\nu^{}_\beta} \hspace{-0.2cm} & = & \hspace{-0.2cm}
\sum_\alpha \left(\Phi^{\rm S}_{\nu^{}_\alpha} P^{}_{\alpha \beta}
\right) \; ,
\nonumber \\
\Phi^{\rm T}_{\overline{\nu}^{}_\beta} \hspace{-0.2cm} & = &
\hspace{-0.2cm} \sum_\alpha \left(\Phi^{\rm
S}_{\overline{\nu}^{}_\alpha} \overline{P}^{}_{\alpha \beta} \right)
\; ,
\end{eqnarray}
where $P^{}_{\alpha\beta} \equiv P(\nu^{}_\alpha \to \nu^{}_\beta)$
and $\overline{P}^{}_{\alpha\beta} \equiv P(\overline{\nu}^{}_\alpha
\to \overline{\nu}^{}_\beta)$ stand respectively for the oscillation
probabilities of UHE cosmic neutrinos and antineutrinos. Since the
galactic distances far exceed the observed solar and atmospheric
neutrino oscillation lengths, $P^{}_{\alpha\beta}$ and
$\overline{P}^{}_{\alpha\beta}$ are actually averaged over many
oscillations and thus become energy-independent:
\begin{eqnarray}
P^{}_{\alpha \beta} = \overline{P}^{}_{\alpha\beta}
= \sum_i \left(|V^{}_{\alpha i}|^2 |V^{}_{\beta i}|^2 \right) \; ,
\end{eqnarray}
where $V^{}_{\alpha i}$ and $V^{}_{\beta i}$ (for $\alpha, \beta =
e, \mu, \tau$ and $i=1,2,3$) denote the elements of the
$3\times 3$ neutrino mixing matrix $V$. For our purpose, we are mainly
interested in the determination of $\Phi^{\rm
T}_{\overline{\nu}^{}_e}$ via the GR channel
$\overline{\nu}^{}_e + e^- \to W^- \to {\rm anything}$. So we
establish a link between three source parameters and a working
observable at the neutrino telescope:
\begin{eqnarray}
R^{}_0 \hspace{-0.2cm} & \equiv & \hspace{-0.2cm}
\frac{\Phi^{\rm T}_{\overline{\nu}^{}_e}}
{\Phi^{\rm T}_{\nu^{}_\mu} + \Phi^{\rm T}_{\overline{\nu}^{}_\mu}}
\nonumber \\
\hspace{-0.2cm} & = & \hspace{-0.2cm}
\left[\frac{1+\delta^{}_{pp}}{2+\delta^{}_{pp}} -
\frac{1+\delta^{}_{pp} -
\delta^{}_{p\gamma}}{(2+\delta^{}_{pp})(1+\delta^{}_{p\gamma})} \kappa
\right] \frac{P^{}_{ee}}{P^{}_{e\mu} + 2 P^{}_{\mu\mu}}
+ \frac{P^{}_{e\mu}}{P^{}_{e\mu} + 2 P^{}_{\mu\mu}} \; ,
\end{eqnarray}
where $P^{}_{ee}$, $P^{}_{e\mu}$ and $P^{}_{\mu\mu}$ can directly
be read off from Eq. (4). After the matrix elements of $V$ are
determined to a sufficiently good degree of accuracy in solar,
atmospheric, reactor and accelerator neutrino oscillation
experiments, a measurement of $R^{}_0$ at a neutrino telescope
will allow one to constrain the source parameters via Eq. (5).
There are two special cases, corresponding
to the pure $p\gamma$ interaction ($\kappa = 1$) and the pure
$pp$ interaction ($\kappa = 0$) at the astrophysical source
of cosmic rays:
\begin{eqnarray}
R^{}_0 (\kappa = 1) \hspace{-0.2cm} & = & \hspace{-0.2cm}
\frac{\delta^{}_{p\gamma}}{1+\delta^{}_{p\gamma}} \cdot
\frac{P^{}_{ee}}{P^{}_{e\mu} + 2 P^{}_{\mu\mu}}
+ \frac{P^{}_{e\mu}}{P^{}_{e\mu} + 2 P^{}_{\mu\mu}} \; , \nonumber \\
R^{}_0 (\kappa =0) \hspace{-0.2cm} & = & \hspace{-0.2cm}
\frac{1+\delta^{}_{pp}}{2+\delta^{}_{pp}} \cdot
\frac{P^{}_{ee}}{P^{}_{e\mu} + 2 P^{}_{\mu\mu}}
+ \frac{P^{}_{e\mu}}{P^{}_{e\mu} + 2 P^{}_{\mu\mu}} \; .
\end{eqnarray}
If both $\delta^{}_{p\gamma}$ and $\delta^{}_{pp}$ are switched
off, then Eq. (5) can be simplified to
\begin{eqnarray}
R^{}_0 (\delta^{}_{p\gamma} = \delta^{}_{pp} =0)
\hspace{-0.2cm} & = & \hspace{-0.2cm}
\frac{1 -\kappa}{2} \cdot \frac{P^{}_{ee}}{P^{}_{e\mu} +
2 P^{}_{\mu\mu}}
+ \frac{P^{}_{e\mu}}{P^{}_{e\mu} + 2 P^{}_{\mu\mu}} \; .
\end{eqnarray}
This result is particularly interesting in the sense that it offers an
opportunity to determine $\kappa$ in a cosmic accelerator
from the measurement of $R^{}_0$ at a neutrino telescope.

In the standard parametrization of $V$ \cite{PDG}, $P^{}_{ee}$,
$P^{}_{e\mu}$ and $P^{}_{\mu\mu}$ can be expressed in terms of three
neutrino mixing angles $(\theta^{}_{12}, \theta^{}_{23},
\theta^{}_{13})$ and the Dirac-type CP-violating phase $\delta$ as
follows:
\begin{eqnarray}
P^{}_{e e} \hspace{-0.2cm} & \simeq & \hspace{-0.2cm}
1 - \frac{1}{2}\sin^2 2\theta^{}_{12} -
\left( 2 - \sin^2 2\theta^{}_{12} \right) \sin^2\theta^{}_{13} \; ,
\nonumber \\
P^{}_{e \mu} \hspace{-0.2cm} & \simeq & \hspace{-0.2cm} \frac{1}{2}
\sin^2 2\theta^{}_{12} \cos^2 \theta^{}_{23} + \frac{1}{4} \sin
4\theta^{}_{12} \sin 2\theta^{}_{23} \sin \theta^{}_{13} \cos \delta
+ \left( 2\sin^2 \theta^{}_{23} - \frac{1}{2} \sin^2 2\theta^{}_{12}
\right) \sin^2\theta^{}_{13} \; ,
\nonumber \\
P^{}_{\mu \mu} \hspace{-0.2cm} & \simeq & \hspace{-0.2cm}
1 - \frac{1}{2} \sin^2 2\theta^{}_{23}
-\frac{1}{2} \sin^2 2\theta^{}_{12} \cos^4 \theta^{}_{23} -
\frac{1}{2} \sin 4 \theta^{}_{12} \sin 2\theta^{}_{23} \cos^2
\theta^{}_{23} \sin \theta^{}_{13} \cos \delta \nonumber \\
\hspace{-0.2cm} && \hspace{-0.2cm}
+ \frac{1}{4} \left[ \sin^2 2\theta^{}_{12} \sin^2 2\theta^{}_{23}
\left( 2 + \cos 2\delta \right) - 8 \sin^4\theta^{}_{23} \right]
\sin^2\theta^{}_{13} \; ,
\end{eqnarray}
in which the terms proportional to $\sin^3\theta^{}_{13} \sim
0.3\%$ and those much smaller ones have been omitted. A global
analysis of the latest neutrino oscillation data \cite{Fogli}
yield $\sin^2\theta^{}_{12} = 0.306^{+0.018}_{-0.015}$,
$\sin^2\theta^{}_{13} = 0.021^{+0.007}_{-0.008}$ and
$\sin^2\theta^{}_{23} = 0.42^{+0.08}_{-0.03}$ at the $1\sigma$ level
\footnote{Note that these results are obtained by using the old
reactor antineutrino fluxes \cite{Fogli}. If the new reactor
antineutrino fluxes \cite{New} are used, the corresponding best-fit values
and 1$\sigma$ ranges of $\sin^2\theta^{}_{12}$ and
$\sin^2\theta^{}_{13}$ will be shifted by about $+0.006$
and $+0.004$, respectively, but the result of $\sin^2\theta^{}_{23}$ is
essentially unchanged \cite{Fogli}.},
while the Dirac-type CP-violating phase $\delta$ remains entirely
unrestricted. Because the contributions of $\delta$ to
$P^{}_{ee}$, $P^{}_{e\mu}$ and $P^{}_{\mu\mu}$ are always
suppressed by small $\sin\theta^{}_{13}$, the $\delta$-induced
uncertainties in the calculation of $R^{}_0$ should not be significant.

Note that a real observable of the GR channel
$\overline{\nu}^{}_e + e^- \to W^- \to {\rm anything}$ at a neutrino
telescope can be the ratio of the $\overline{\nu}^{}_e$ events to the
$\nu^{}_\mu$ and $\overline{\nu}^{}_\mu$ events of charged-current
interactions in the vicinity of the resonance $E^{}_{\overline{\nu}^{}_e}
\simeq M^2_W/(2 m^{}_e) \simeq 6.3$ PeV \cite{Indian,Xing06}:
\begin{eqnarray}
R \equiv \frac{N^{}_{\overline{\nu}^{}_e}}{N^{\rm CC}_{\nu^{}_\mu}
+ N^{\rm CC}_{\overline{\nu}^{}_\mu}} = a R^{}_0 \; ,
\end{eqnarray}
where $a \simeq 30.5$ can be obtained in an optimal case by
assuming the $E^{-2}_{\nu^{}_\alpha}$ neutrino spectrum
\cite{Indian} and considering the muon events with contained
vertices \cite{Beacom} in a water- or ice-based detector. A more
accurate calculation of $a$ is certainly crucial for the IceCube
detector to detect the rate of the GR reaction
\cite{Weiler}. Note also that the $\overline{\nu}^{}_e$ flux of
$E^{}_{\overline{\nu}^{}_e} \simeq 6.3$ PeV might largely get
absorbed in passing through the Earth \cite{Indian}. Hence it is
only feasible for a neutrino telescope to detect the downward-going
or horizontal $\overline{\nu}^{}_e$ flux whose energy lies in the
vicinity of the GR, in which case the atmospheric
neutrino flux of the same energy is negligibly small and should not
be of concern as an important background \cite{Indian}.

We proceed to illustrate the dependence of $R^{}_0$ on $\kappa$,
$\delta^{}_{p\gamma}$ and $\delta^{}_{pp}$ with the help of current
experimental data on three neutrino mixing angles. First of all, we
assume $\delta^{}_{p\gamma} = \delta^{}_{pp} =0$ and use Eq. (7) to
describe the relationship between $R^{}_0$ and $\kappa$. Fig. 1
shows the allowed region of $R^{}_0$ versus $0 \leq \kappa \leq 1$,
where the $1\sigma$ ranges of $\theta^{}_{12}$, $\theta^{}_{13}$ and
$\theta^{}_{23}$ together with $\delta \in [0,2\pi)$ have been
scanned. The central value of $R^{}_0$ for a given value of $\kappa$
is calculated by inputting the best-fit values of three neutrino
mixing angles (i.e., $\sin^2 \theta_{12} = 0.306$, $\sin^2
\theta_{13} = 0.021$ and $\sin^2 \theta_{23} = 0.42$ \cite{Fogli})
and taking $\delta = 0$. Although the uncertainties associated with
four neutrino mixing parameters remain rather large, we have the
following quantitative observations: (1) the magnitude of $R^{}_0$
is restricted to the range $0.18 \leq R^{}_0 \leq 0.58$; (2)
$R^{}_0$ lies in the range $0.18 \leq R^{}_0 \leq 0.31$ for the pure
$p\gamma$ interaction (i.e., $\kappa = 1$); and (3) $R^{}_0$ lies in
the range $0.45 \leq R^{}_0 \leq 0.58$ for the pure $pp$ interaction
(i.e., $\kappa = 0$). As the neutrino mixing parameters can be more
and more precisely measured in the ongoing and future neutrino
oscillation experiments, we expect that the GR will serve as a clear
discriminator of UHE cosmic neutrinos originating from $pp$ and
$p\gamma$ collisions at an astrophysical source.

Now let us examine possible effects of $\delta^{}_{p\gamma}$ and
$\delta^{}_{pp}$ on the relationship between $R^{}_0$ and $\kappa$.
For simplicity, we only take the best-fit values of three neutrino
mixing angles and assume $\delta =0$ in our numerical illustration.
The change of $R^{}_0$ with respect to three source parameters
$\kappa$, $\delta^{}_{p\gamma}$ and $\delta^{}_{pp}$ is shown in
Fig. 2, where $\delta_{p\gamma} \in [0, \ +0.2]$ and $\delta_{pp}
\in [-0.2, \ +0.2]$ have been assumed. Note that
$\delta^{}_{p\gamma}$ is positive (or vanishing) by definition,
while $\delta^{}_{pp}$ can be either positive or negative (or
vanishing), corresponding to an excess of the $\pi^-$ or $\pi^+$
events (or $\Phi^{pp}_{\pi^+} = \Phi^{pp}_{\pi^-}$) in the $pp$
interaction at an astrophysical source. As in Fig. 1, the central
curve of $R^{}_0$ varying with $\kappa$ in Fig. 2 is obtained in the
assumption of $\delta^{}_{p\gamma} = \delta^{}_{pp} = 0$. It is
straightforward to see that $\delta^{}_{p\gamma}$ and
$\delta^{}_{pp}$ can significantly affect $R^{}_0$ for a given value
of $\kappa$. For the pure $p\gamma$ interaction with $\kappa = 1$, a
variation of $\delta^{}_{p\gamma}$ from 0 to $0.2$ results in a
change of $R^{}_0$ by more than $30\%$ as compared with its original
value. As indicated by Eq. (6), it is in principle possible to
determine or constrain the free parameter $\delta^{}_{p\gamma}$ (or
$\delta^{}_{pp}$) for a given source with the pure $p\gamma$ (or
$pp$) interaction by measuring $R^{}_0$ at a neutrino telescope.

If the uncertainties from both the neutrino mixing parameters
($\theta^{}_{12}$, $\theta^{}_{13}$, $\theta^{}_{23}$ and $\delta$)
and the source parameters ($\delta^{}_{p\gamma}$ and
$\delta^{}_{pp}$) are taken into account, it will be almost
impossible to distinguish between $p\gamma$ and $pp$ collisions even
if $R^{}_0 \sim 0.4$ is extracted from a neutrino telescope
experiment. This observation implies that it {\it does} make sense
for us to consider the nontrivial effects of $\delta^{}_{p\gamma}$
and $\delta^{}_{pp}$. What we can do at present is to carefully
study the yields of $\pi^\pm$ fluxes in the realistic models of
$p\gamma$ and $pp$ collisions, so as to obtain some theoretical
constraints on $\delta^{}_{p\gamma}$ and $\delta^{}_{pp}$
\cite{Winter}. In addition, we must determine the neutrino mixing
parameters as precisely as possible in all the terrestrial neutrino
oscillation experiments.

\section{Estimate of the Event Rate and Background}

To further illustrate, let us estimate the event rate of
the GR signal and the relevant background. We assume the total
flux of UHE cosmic neutrinos and antineutrinos originating
from an optically thin source
to saturate the Waxman-Bahcall (WB) bound \cite{WB}
\begin{equation}
E^2_\nu \Phi^{}_{\nu+\overline{\nu}} = 2\times 10^{-8}
~\epsilon^{}_\pi~\xi^{}_z~{\rm GeV}~{\rm cm}^{-2}~{\rm s}^{-1}~{\rm
sr}^{-1} \; ,
\end{equation}
where $\epsilon^{}_\pi$ stands for the ratio of the pion energy to the
initial proton energy, and $\xi^{}_z \approx 3$ for a source
evolution $\propto (1+z)^3$ with $z$ being the redshift. We have
$\epsilon^{}_\pi = \epsilon^{p\gamma}_\pi \approx 0.25$ for
$p\gamma$ collisions or $\epsilon^{}_\pi = \epsilon^{pp}_\pi
\approx 0.6$ for $pp$ collisions. Therefore, the WB bound actually
depends on whether the $pp$ or $p\gamma$ collision is assumed. Since
there is on average one cosmic-ray neutron produced per proton
collision, we may parametrize $\Phi^{}_{\nu + \overline{\nu}}$
saturating the WB bound as
\begin{equation}
E^2_\nu \Phi^{}_{\nu+\overline{\nu}} = 6 \times 10^{-8} \left[\left(1 -
\kappa^\prime\right) \epsilon^{pp}_\pi + \kappa^\prime
\epsilon^{p\gamma}_\pi \right]~{\rm GeV}~{\rm cm}^{-2}~{\rm
s}^{-1}~{\rm sr}^{-1} \; ,
\end{equation}
where $\kappa^\prime$ denotes the fraction of the $p\gamma$
collisions. In this parametrization $\kappa^\prime = 1$ and
$\kappa^\prime = 0$ correspond to the pure $p\gamma$ and pure $pp$
interactions, respectively. Note that we have defined $\kappa$ in
Eq. (1) as the fraction of the pion fluxes from the $p\gamma$ collisions.
The relationship between $\kappa$ and $\kappa^\prime$ can be easily
established:
\begin{equation}
\kappa^\prime = \frac{\kappa \ \epsilon^{pp}_\pi}{\left(1 -
\kappa\right)\epsilon^{p\gamma}_\pi + \kappa \ \epsilon^{pp}_\pi} \; .
\end{equation}
Given the total flux of neutrinos and antineutrinos in Eq. (11) and
their flavor distribution at the source in Eq. (2), it is then possible
to calculate the neutrino and antineutrino fluxes of different
flavors at a neutrino telescope by taking account of the effect
of flavor oscillations. We obtain
\begin{eqnarray}
\Phi^{\rm T}_{\nu^{}_\alpha} &=& \Phi^{}_0 \frac{\epsilon^{pp}_\pi
\ \epsilon^{p\gamma}_\pi}{\left(1-\kappa\right) \epsilon^{p\gamma}_\pi + \kappa
\ \epsilon^{pp}_\pi} \times \left\{\frac{1}{3}
\left[\frac{1}{2+\delta^{}_{pp}} + \frac{1+\delta^{}_{pp} -
\delta^{}_{p\gamma}}{(2+\delta^{}_{pp})
(1+\delta^{}_{p\gamma})}\kappa \right]P^{}_{e\alpha} + \frac{1}{3}
P^{}_{\mu \alpha}\right\} \; ,
\nonumber \\
\Phi^{\rm T}_{\overline{\nu}^{}_\alpha} &=& \Phi^{}_0
\frac{\epsilon^{pp}_\pi \ \epsilon^{p\gamma}_\pi}{\left(1-\kappa\right)
\epsilon^{p\gamma}_\pi + \kappa \ \epsilon^{pp}_\pi} \times
\left\{\frac{1}{3} \left[\frac{1+\delta^{}_{pp}}{2+\delta^{}_{pp}} -
\frac{1+\delta^{}_{pp} - \delta^{}_{p\gamma}}{(2+\delta^{}_{pp})
(1+\delta^{}_{p\gamma})} \kappa \right]P^{}_{e\alpha} + \frac{1}{3}
P^{}_{\mu \alpha}\right\} \; , ~~
\end{eqnarray}
where Eq. (3) has been used and $\Phi^{}_0 \equiv 6\times
10^{-8}~{\rm GeV}^{-1}~{\rm cm}^{-2}~{\rm s}^{-1}~{\rm sr}^{-1}
\left(1~{\rm GeV}/E^{}_\nu\right)^2$ is defined. Note that the energy
dependence of $\nu^{}_\alpha$ and $\overline{\nu}^{}_\alpha$ fluxes
in Eq. (13) has been suppressed.

Following Ref. \cite{Weiler}, we estimate the event rate of the GR
signal in the IceCube experiment:
\begin{equation}
{\rm d}N^{}_{\rm s}/{\rm d}t = 66.7\% \times  N^{}_{\rm eff} \Delta
\Omega \int {\rm d}E^{}_\nu \Phi^{\rm
T}_{\overline{\nu}^{}_e}(E^{}_\nu) \sigma^{}_{\rm GR}(E^{}_\nu) \; ,
\end{equation}
in which the coefficient $66.7\%$ is the branching ratio of hadronic
$W^-$ decays, $N^{}_{\rm eff} \approx 6\times 10^{38}$ denotes the
number of target electrons for an effective volume $V^{}_{\rm eff}
\sim 2~{\rm km}^3$ of the IceCube detector, $\Delta \Omega \approx
2\pi$ is the solid angle aperture, and $\sigma^{}_{\rm GR}(E^{}_\nu)
= \pi g^2 M^2_W \delta(2m^{}_e E^{}_\nu - M^2_W)/(4m^{}_e E^{}_\nu)$
is the cross section of the GR scattering. The typical GR signal is
the shower events induced by the hadronic decays of $W^-$ in the
resonant energy region, while the main background comes from the
non-resonant inelastic scattering of $\nu^{}_e$ and
$\overline{\nu}^{}_e$ with nucleons in the detector. As for the
background events, the effective number of target nucleons is
approximately twice the number of electrons (i.e., $N^\prime_{\rm
eff} \approx 1.2\times 10^{39}$) and the solid angle aperture is
$\Delta \Omega^\prime \approx 4\pi$. The cross sections of
charged-current $\nu^{}_e N$ and $\overline{\nu}^{}_e N$
interactions are well represented by the power-law forms
\cite{Gandhi}:
\begin{eqnarray}
\sigma^{\nu N}_{\rm CC}(E^{}_\nu) &=& 2.69\times 10^{-36}~{\rm cm}^2
\left(\frac{E^{}_\nu}{1~{\rm GeV}}\right)^{0.402} \;, \nonumber \\
\sigma^{\overline{\nu} N}_{\rm CC}(E^{}_\nu) &=& 2.53\times
10^{-36}~{\rm cm}^2 \left(\frac{E^{}_\nu}{1~{\rm
GeV}}\right)^{0.404} \;.
\end{eqnarray}
Integrating over the resonant acceptance energy bin $(10^{6.7}\cdots
10^{6.9})~{\rm GeV}$ for the IceCube telescope, we can obtain the event
rate for the background
\begin{equation}
{\rm d}N^{}_{\rm b}/{\rm d}t = N^\prime_{\rm eff} \Delta
\Omega^\prime \int^{10^{6.9} {\rm GeV}}_{10^{6.7} {\rm GeV}} {\rm
d}E^{}_\nu \left[\Phi^{\rm T}_{\nu^{}_e}(E^{}_\nu) \sigma^{\nu
N}_{\rm CC}(E^{}_\nu) + \Phi^{\rm T}_{\overline{\nu}^{}_e}(E^{}_\nu)
\sigma^{\overline{\nu} N}_{\rm CC}(E^{}_\nu)\right] \; .
\end{equation}
As usual, the signal-to-background ratio can be defined as $R^{}_{\rm
s/b} \equiv ({\rm d}N^{}_{\rm s}/{\rm d}t)/({\rm d}N^{}_{\rm b}/{\rm
d}t)$, which measures the significance of the signal events.

We perform a numerical calculation of the event rate of the GR
signal ${\rm d}N^{}_{\rm s}/{\rm d}t$ and the signal-to-background
ratio $R^{}_{{\rm s}/{\rm b}}$, and examine their dependence on the
source parameters $(\kappa, \delta^{}_{p\gamma}, \delta^{}_{pp})$
and the neutrino mixing parameters $(\theta^{}_{12}, \theta^{}_{23},
\theta^{}_{13}, \delta)$. Fig. 3 shows the expected event rate ${\rm
d}N^{}_{\rm s}/{\rm d}t$ versus $0\leq \kappa \leq 1$, where
$\delta^{}_{p\gamma} = \delta^{}_{pp} = 0$ is assumed and the
$1\sigma$ ranges of $\theta^{}_{12}$, $\theta^{}_{13}$ and
$\theta^{}_{23}$ together with $\delta \in [0,2\pi)$ have been
scanned. The corresponding signal-to-background ratio in this case
is shown in Fig. 4. For the pure $pp$ interaction (i.e., $\kappa =
0$), we obtain ${\rm d}N^{}_{\rm s}/{\rm d}t \approx 3.5$ per year
and $R^{}_{{\rm s}/{\rm b}} \approx 5$, indicating a great discovery
potential of the IceCube telescope after several years of data
accumulation \cite{Weiler}. For the pure $p\gamma$ interaction
(i.e., $\kappa = 1$), however, the event rate is quite low: ${\rm
d}N^{}_{\rm s}/{\rm d}t \approx 0.8$ per year. Hence it is quite
challenging for the IceCube detector to discover UHE cosmic
neutrinos originating from an optically thin source with the pure
$p\gamma$ collisions. This observation justifies the importance of
the GR channel in distinguishing between $pp$ and $p\gamma$
interactions. On the other hand, the dependence of ${\rm d}N^{}_{\rm
s}/{\rm d}t$ and $R^{}_{{\rm s}/{\rm b}}$ on the source parameters
$(\kappa, \delta^{}_{p\gamma}, \delta^{}_{pp})$ are illustrated in
Fig. 5 and Fig. 6, where the best-fit values of three neutrino
mixing angles (i.e., $\sin^2 \theta_{12} = 0.306$, $\sin^2
\theta_{13} = 0.021$ and $\sin^2 \theta_{23} = 0.42$ \cite{Fogli})
together with $\delta = 0$ have been input. It is straightforward to
see the degeneracy between the uncertainty induced by those neutrino
mixing parameters and that by the source parameters
$(\delta^{}_{p\gamma}, \delta^{}_{pp})$. So a full determination of
the latter requires more precise values of neutrino oscillation
parameters and a neutrino telescope whose scale should be much
larger than the IceCube detector.

\section{Summary}

We have re-examined the possibility of using the GR
channel $\overline{\nu}^{}_e + e^- \to W^- \to {\rm
anything}$ to discriminate between the UHE cosmic neutrinos
originating from $p\gamma$ and $pp$ collisions in an optically thin
source of cosmic rays. After proposing a general parametrization of
the initial neutrino flavor distribution by taking account of
non-zero $\delta^{}_{p\gamma}$ and $\delta^{}_{pp}$ at the source,
we have established an analytical relationship between the typical
source parameter $\kappa$ and the working observable of the GR
$R^{}_0$ at a neutrino telescope. We have also illustrated
the numerical dependence of $R^{}_0$ on $\kappa$ with the help of
the latest experimental data on three neutrino mixing angles. We
find that a measurement of $R^{}_0$ is in principle possible to
identify the pure $p\gamma$ interaction ($\kappa =1$), the pure $pp$
interaction ($\kappa =0$) or a mixture of both of them ($0 < \kappa
< 1$) at a given source of UHE cosmic neutrinos.
In addition, the event rate of the GR signal against the
relevant background is estimated by assuming the total flux of
UHE cosmic neutrinos and antineutrinos originating from an optically
thin source to saturate the WB bound.

A measurement of the GR and a determination of the
flavor distribution of UHE cosmic neutrinos at an astrophysical
source are certainly big challenges
to the IceCube detector and other possible neutrino telescopes.
Anyway, our present understanding of the production mechanism of
UHE cosmic neutrinos depends on a number of hypotheses and thus needs
more and more observational supports. We therefore expect that neutrino
telescopes can help us in this connection in the long run.

\vspace{0.5cm}

We would like to thank S. Pakvasa and W. Winter for their useful
comments and discussions.
This work was supported in part by the National Natural
Science Foundation of China under grant No. 10875131 (Z.Z.X.) and
by the Alexander von Humboldt Foundation (S.Z.).

\newpage

\begin{figure}
\vspace{1.5cm}
\epsfig{file=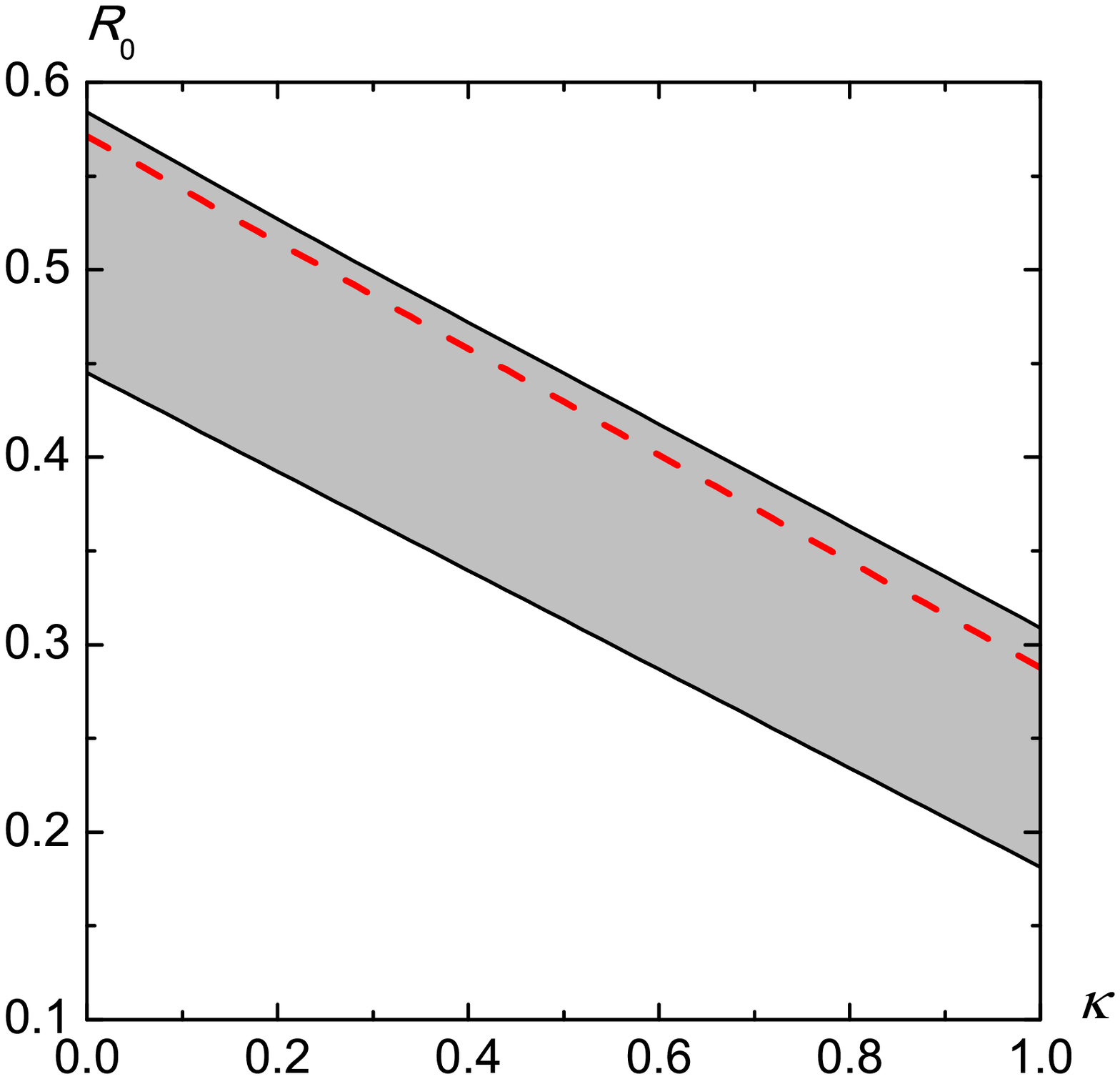,bbllx=-5cm,bblly=10.5cm,bburx=0cm,bbury=15.5cm,%
width=2.5cm,height=2.5cm,angle=0,clip=0}\vspace{4.2cm} \caption{The
dependence of the working observable $R^{}_0$ on the source
parameter $\kappa$ in the assumption of $\delta_{p\gamma} =
\delta_{pp} = 0$. The dashed curve corresponds to the best-fit
values of $\theta^{}_{12}$, $\theta^{}_{13}$ and $\theta^{}_{23}$
together with $\delta = 0$, and the uncertainties come from the
$1\sigma$ error bars of three neutrino mixing angles and an
arbitrary change of $\delta \in [0, 2\pi)$.}
\end{figure}

\begin{figure}
\vspace{1.5cm}
\epsfig{file=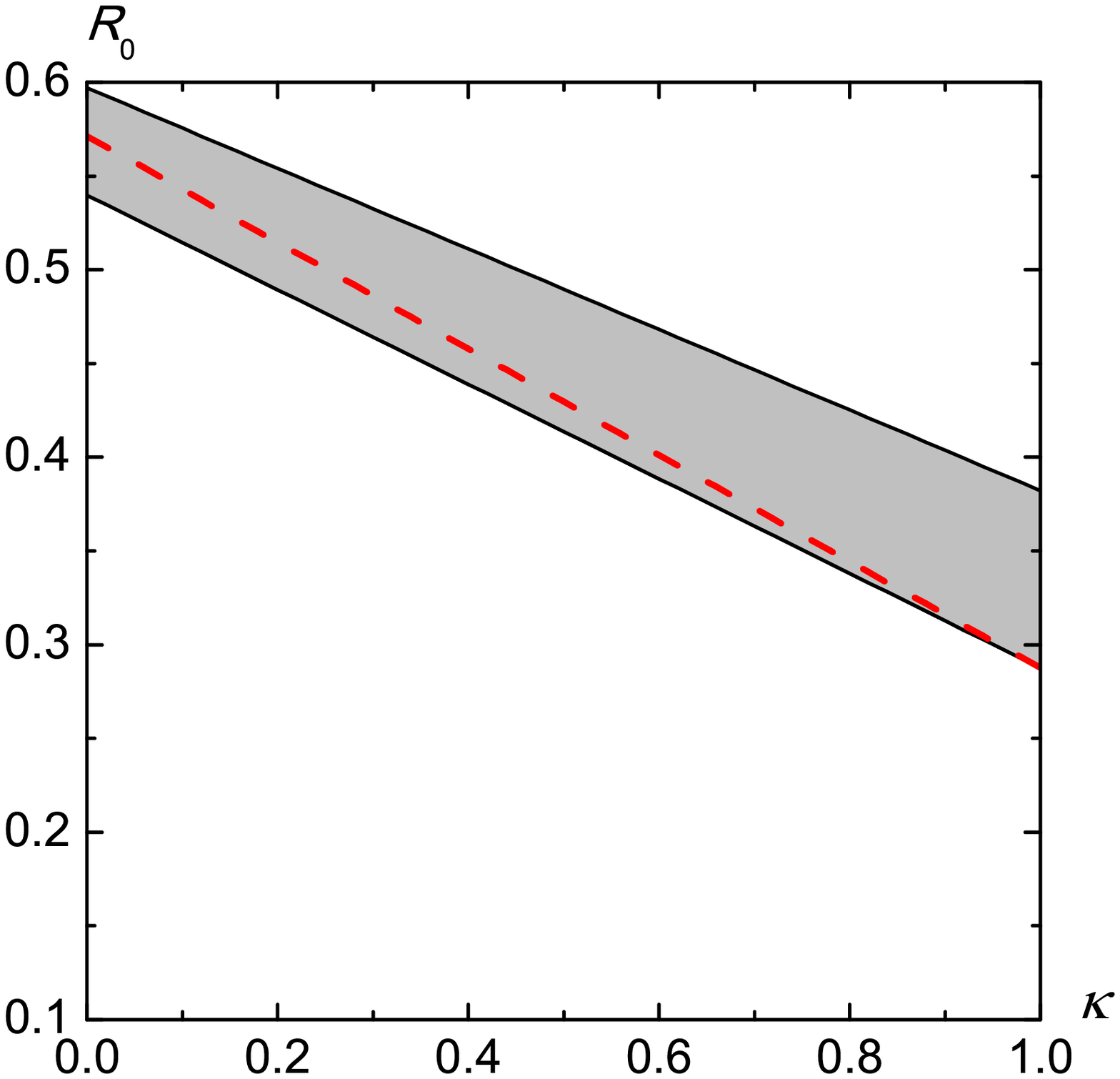,bbllx=-5cm,bblly=11cm,bburx=0cm,bbury=16cm,%
width=2.5cm,height=2.5cm,angle=0,clip=0}\vspace{4.5cm} \caption{The
dependence of the working observable $R^{}_0$ on the source
parameter $\kappa$, where the best-fit values of $\theta^{}_{12}$,
$\theta^{}_{13}$ and $\theta^{}_{23}$ together with $\delta = 0$
have been taken. The dashed curve corresponds to $\delta_{p\gamma}
= \delta_{pp} = 0$, and the uncertainties come from the variations
of $\delta^{}_{p\gamma}$ and $\delta^{}_{pp}$ in the ranges
$\delta_{p\gamma} \in [0, \ +0.2]$ and $\delta_{pp} \in [-0.2, \ +0.2]$.}
\end{figure}

\begin{figure}
\vspace{1.5cm}
\epsfig{file=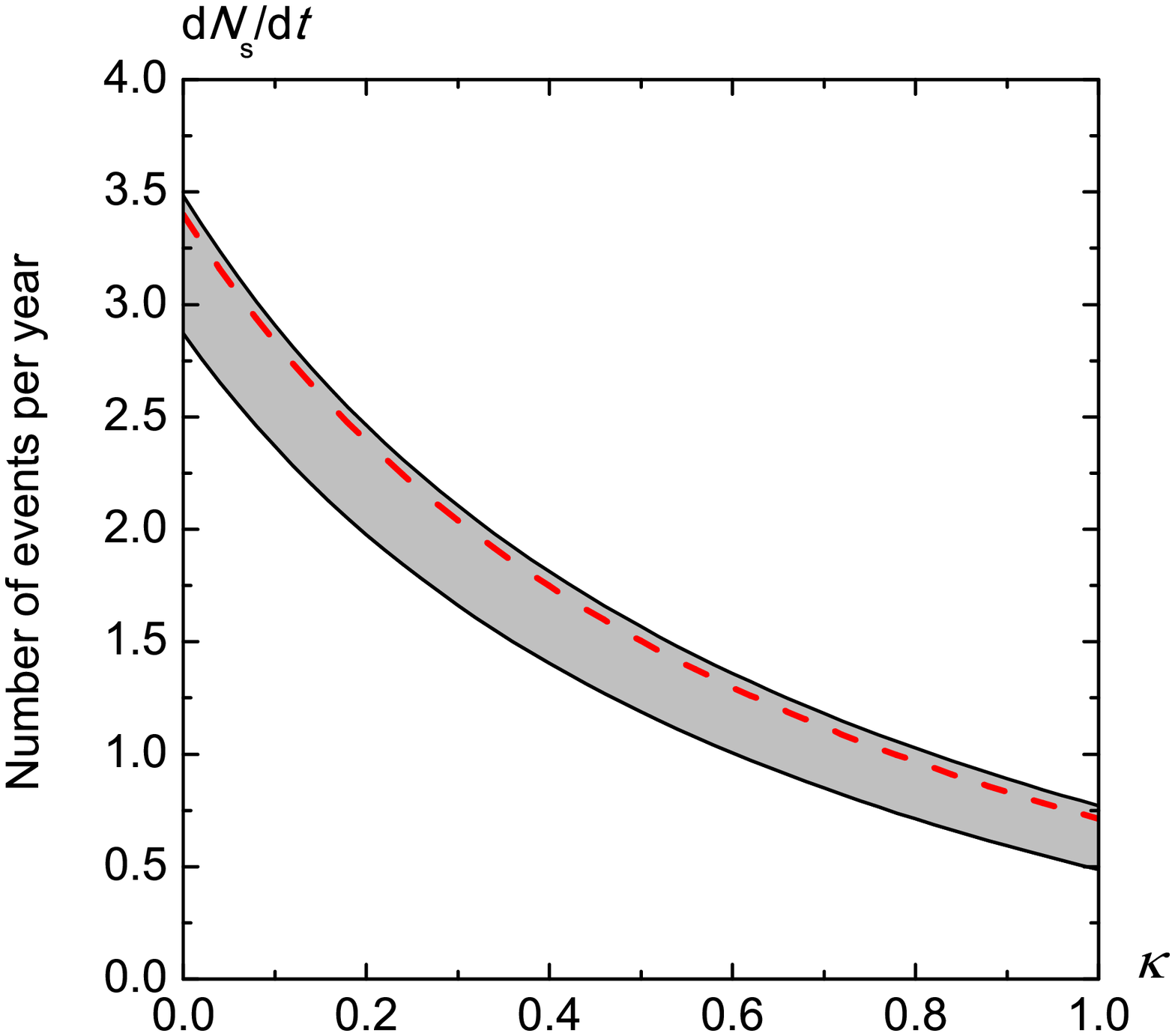,bbllx=-7cm,bblly=10.5cm,bburx=-2cm,bbury=15.5cm,%
width=2.5cm,height=2.5cm,angle=0,clip=0}\vspace{4.2cm} \caption{The
dependence of the event rate of the GR signal ${\rm d}N^{}_{\rm
s}/{\rm d}t$ on the source parameter $\kappa$ in the assumption of
$\delta_{p\gamma} = \delta_{pp} = 0$. The dashed curve corresponds
to the best-fit values of $\theta^{}_{12}$, $\theta^{}_{13}$ and
$\theta^{}_{23}$ together with $\delta = 0$, and the uncertainties
come from the $1\sigma$ error bars of three neutrino mixing angles
and an arbitrary change of $\delta \in [0, 2\pi)$.}
\end{figure}

\begin{figure}
\vspace{1.5cm}
\epsfig{file=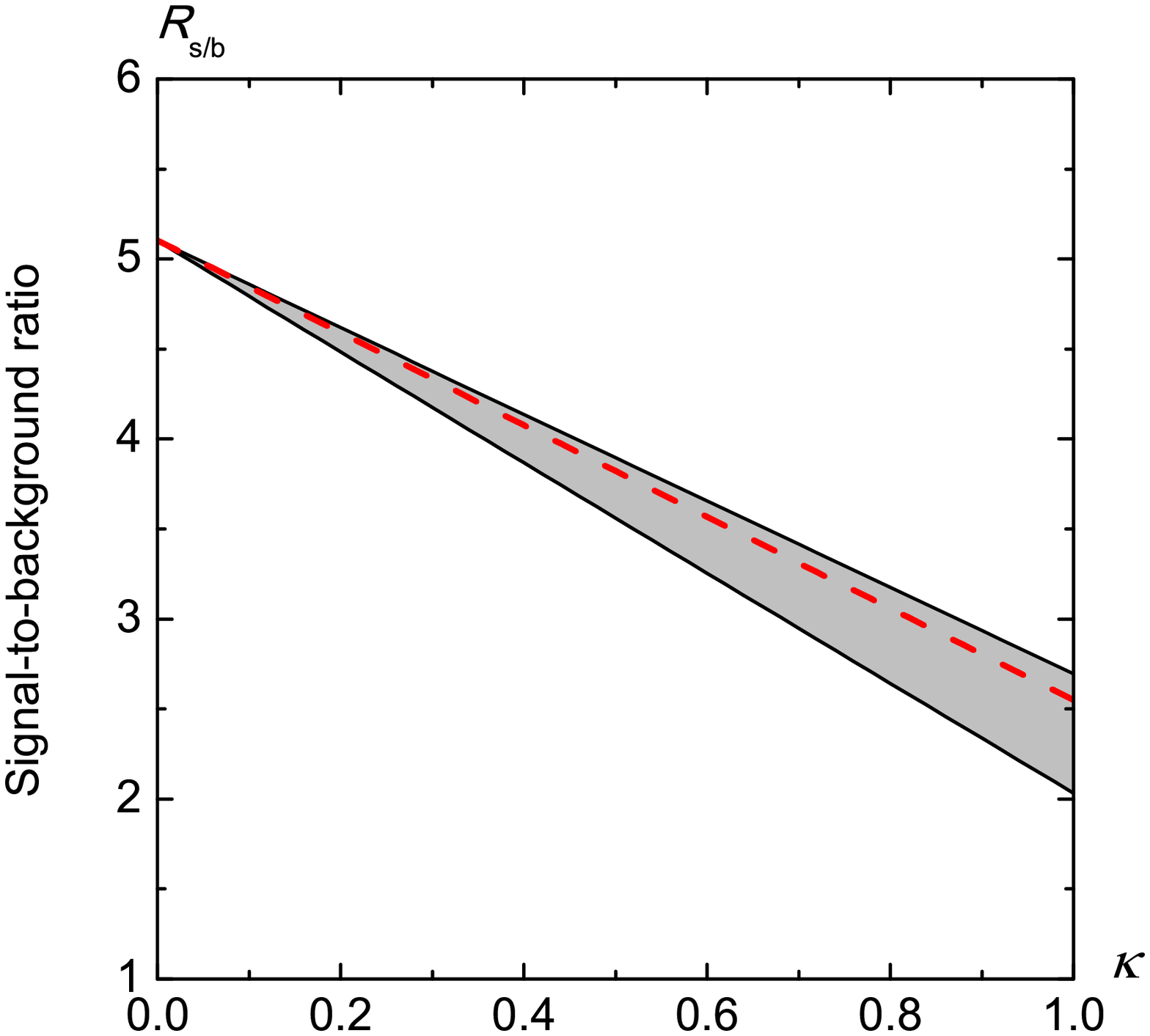,bbllx=-7cm,bblly=11cm,bburx=-2cm,bbury=16cm,%
width=2.5cm,height=2.5cm,angle=0,clip=0}\vspace{4.5cm} \caption{The
dependence of the signal-to-background ratio $R^{}_{\rm s/b}$ on the
source parameter $\kappa$ in the assumption of $\delta_{p\gamma} =
\delta_{pp} = 0$. The dashed curve corresponds to the best-fit
values of $\theta^{}_{12}$, $\theta^{}_{13}$ and $\theta^{}_{23}$
together with $\delta = 0$, and the uncertainties come from the
$1\sigma$ error bars of three neutrino mixing angles and an
arbitrary change of $\delta \in [0, 2\pi)$.}
\end{figure}

\newpage

\begin{figure}
\vspace{1.5cm}
\epsfig{file=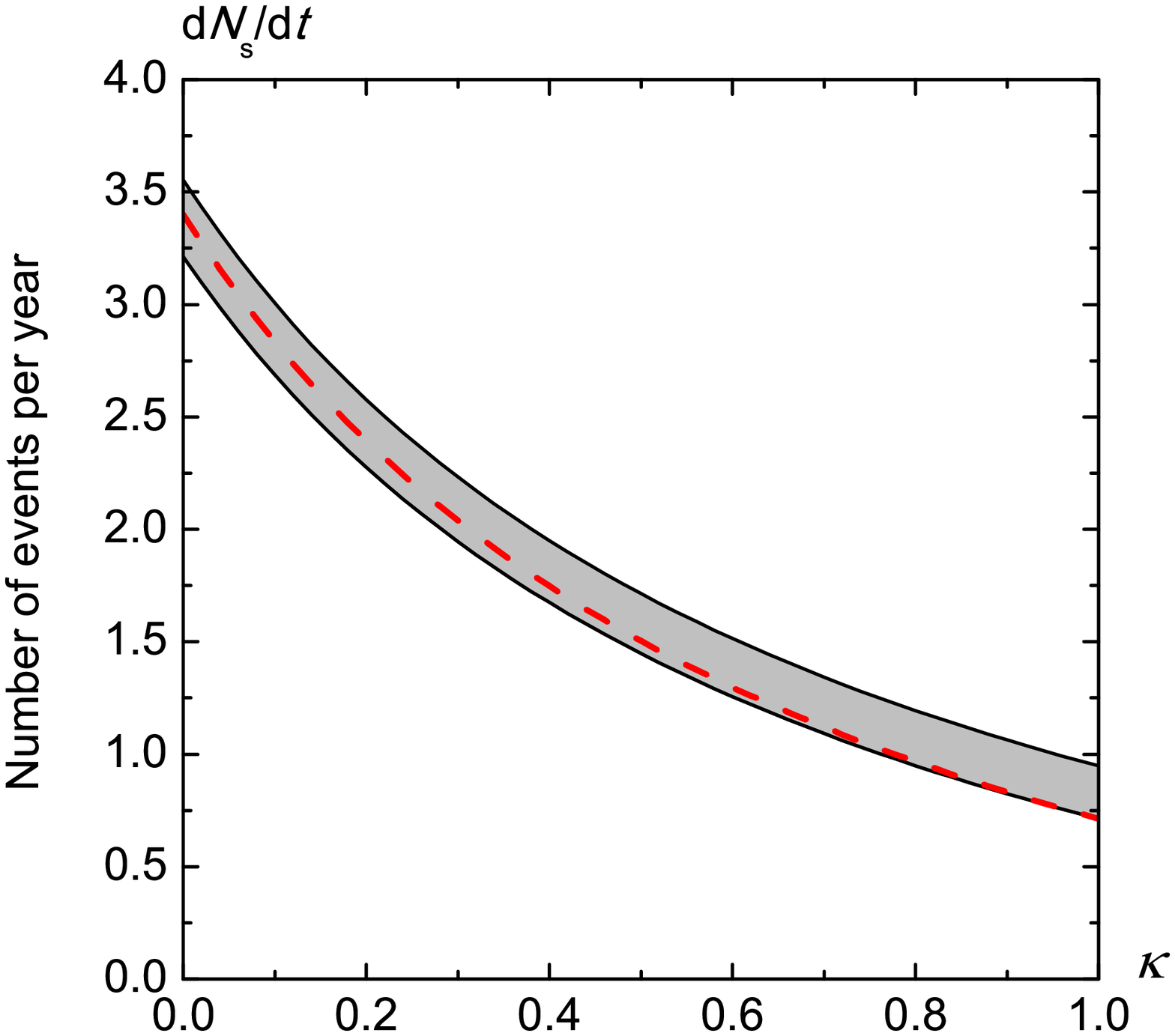,bbllx=-7cm,bblly=10.5cm,bburx=-2cm,bbury=15.5cm,%
width=2.5cm,height=2.5cm,angle=0,clip=0}\vspace{4.2cm} \caption{The
dependence of the event rate of the GR signal ${\rm d}N^{}_{\rm
s}/{\rm d}t$ on the source parameter $\kappa$, where the best-fit
values of $\theta^{}_{12}$, $\theta^{}_{13}$ and $\theta^{}_{23}$
together with $\delta = 0$ have been taken. The dashed curve
corresponds to $\delta_{p\gamma} = \delta_{pp} = 0$, and the
uncertainties come from the variations of $\delta^{}_{p\gamma}$ and
$\delta^{}_{pp}$ in the ranges $\delta_{p\gamma} \in [0, \ +0.2]$
and $\delta_{pp} \in [-0.2, \ +0.2]$.}
\end{figure}

\begin{figure}
\vspace{1.5cm}
\epsfig{file=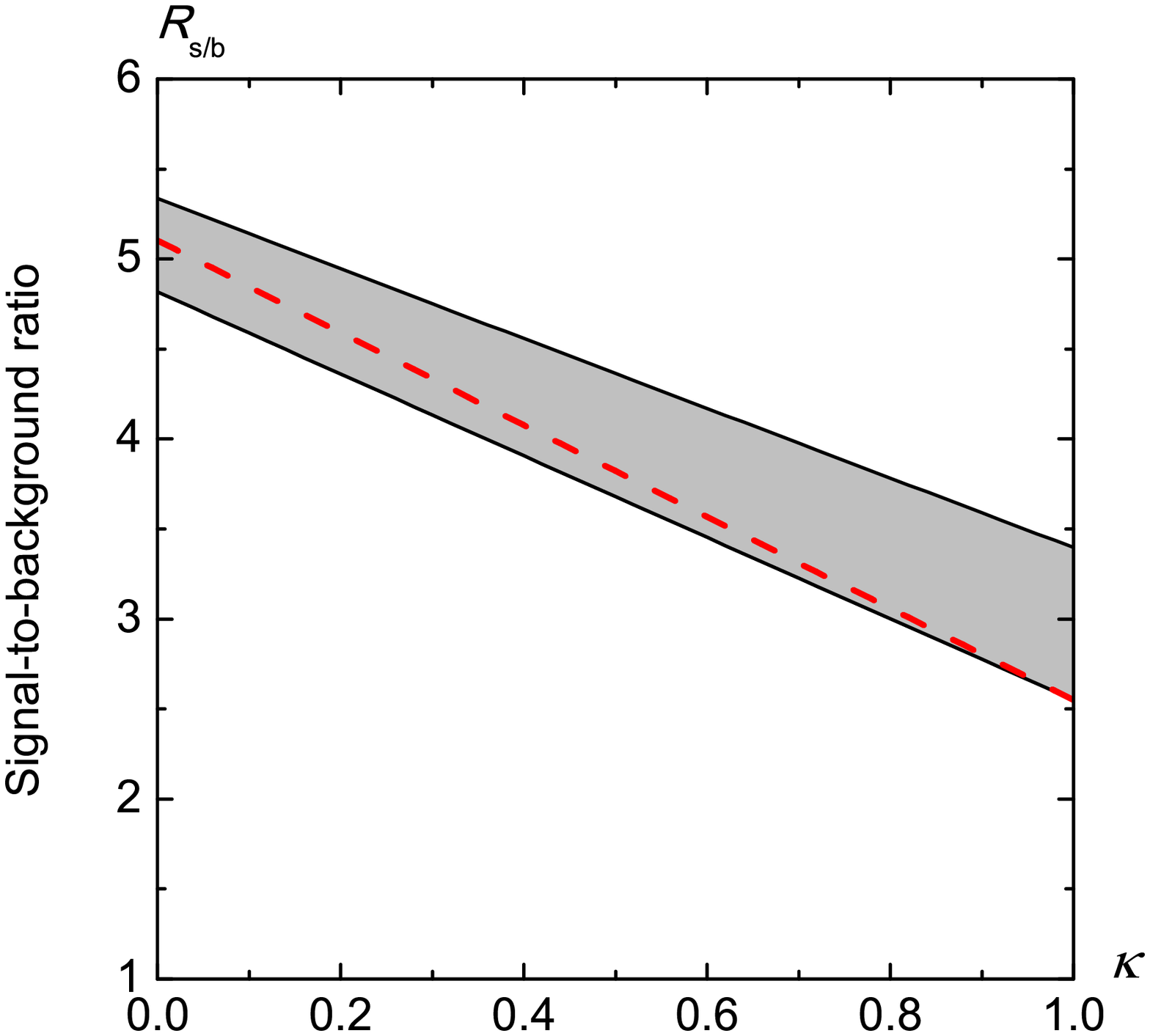,bbllx=-7cm,bblly=11cm,bburx=-2cm,bbury=16cm,%
width=2.5cm,height=2.5cm,angle=0,clip=0}\vspace{4.5cm} \caption{The
dependence of the signal-to-background ratio $R^{}_{\rm s/b}$ on the
source parameter $\kappa$, where the best-fit values of
$\theta^{}_{12}$, $\theta^{}_{13}$ and $\theta^{}_{23}$ together
with $\delta = 0$ have been taken. The dashed curve corresponds to
$\delta_{p\gamma} = \delta_{pp} = 0$, and the uncertainties come
from the variations of $\delta^{}_{p\gamma}$ and $\delta^{}_{pp}$ in
the ranges $\delta_{p\gamma} \in [0, \ +0.2]$ and $\delta_{pp} \in
[-0.2, \ +0.2]$.}
\end{figure}

\newpage

\begin{thebibliography}{99}

\bibitem{IceCube} IceCube Collaboration, J. Ahrens {\it et al.},
Nucl. Phys. Proc. Suppl. {\bf 118}, 388 (2003).

\bibitem{Halzen} For a review with extensive references, see:
F. Halzen and D. Hooper, Rep. Prog. Phys. {\bf 65}, 1025 (2002).

\bibitem{XZ} See, e.g., Z.Z. Xing and S. Zhou, {\it Neutrinos in
Particle Physics, Astronomy and Cosmology} (Zhejiang University
Press and Springer Verlag, 2011).

\bibitem{Pakvasa} J.G. Learned and S. Pakvasa, Astropart. Phys.
{\bf 3}, 267 (1995).

\bibitem{XZ08} Z.Z. Xing and S. Zhou, Phys. Lett. B {\bf 666}, 166
(2008).

\bibitem{Glashow} S.L. Glashow, Phys. Rev. {\bf 118}, 316 (1960).

\bibitem{Berezinsky} V.S. Berezinsky and A.Z. Gazizov, JETP Lett.
{\bf 25}, 254 (1977).


\bibitem{Gandhi} R. Gandhi, C. Quigg, M.H. Reno, and I. Sarcevic,
Astropart. Phys. {\bf 5}, 81 (1996); Phys. Rev. D {\bf 58}, 093009
(1998).

\bibitem{Weiler} L.A. Anchordoqui, H. Goldberg, F. Halzen, and T.J.
Weiler, Phys. Lett. B {\bf 621}, 18 (2005).

\bibitem{Indian} P. Bhattacharjee and N. Gupta, arXiv:hep-ph/0501191.

\bibitem{Winter1} M. Maltoni and W. Winter,
JHEP {\bf 0807}, 064 (2008).

\bibitem{Winter2} S. H$\rm\ddot{u}$mmer, M. Maltoni, W. Winter,
and C. Yaguna, Astropart. Phys. {\bf 34}, 205 (2010).

\bibitem{Lin} The possibility of detecting the UHE cosmic
$\overline{\nu}^{}_e$ flux by means of the $\overline{\nu}^{}_e
+ e^- \to W^- + \gamma$ channel has been discussed in:
H. Athar and G.L. Lin, Astropart. Phys. {\bf 19}, 569 (2003).

\bibitem{Glashow2} S.R. Coleman and S.L. Glashow, arXiv:hep-ph/9808446.

\bibitem{PDG} Particle Data Group, K. Nakamura {\it et al.},
J. Phys. G {\bf 37}, 075021 (2010).

\bibitem{Fogli} G.L. Fogli {\it et al.}, arXiv:1106.6028, in which
the recent indications of $\nu^{}_\mu \to \nu^{}_e$ appearance in the
T2K and MINOS neutrino oscillation experiments have been taken into
account.

\bibitem{New} T.A. Mueller {\it et al.}, Phys. Rev. C {\bf 83},
054615 (2011).

\bibitem{Beacom} J.F. Beacom, N.F. Bell, D. Hooper, S. Pakvasa,
and T.J. Weiler, Phys. Rev. D {\bf 68}, 093005 (2003);
{\bf 72}, 019901(E) (2005).

\bibitem{Xing06} Z.Z. Xing, Phys. Rev. D {\bf 74}, 013009 (2006);
Z.Z. Xing and S. Zhou, Phys. Rev. D {\bf 74}, 013010 (2006).

\bibitem{Winter} S. Hummer, M. Ruger, F. Spanier, and W. Winter,
Astrophys. J. {\bf 721}, 630 (2010).

\bibitem{WB} E. Waxman and J. Bahcall, Phys. Rev. D {\bf 59},
023002 (1999).
\end{thebibliography}
\end{document}